\documentclass[12pt]{article}

\newcommand{\be}{\begin{equation}}
\newcommand{\ee}{\end{equation}}
\newcommand{\bea}{\begin{align}}
\newcommand{\eea}{\end{align}}

\def\({\left(}
\def\){\right)}

\usepackage{color}
\usepackage{multirow}
\usepackage{graphicx}
\usepackage{hyperref}
\usepackage{amsmath}
\usepackage{amssymb}
\usepackage{amsfonts}
\usepackage{latexsym}
\usepackage{slashed}
\usepackage{yfonts}
\usepackage{verbatim}
\usepackage{wrapfig}

\begin{document}

\centerline{\large{Black holes and Bhargava's invariant theory}}
\bigskip
\bigskip
\centerline{Murat Gunaydin$^1$, Shamit Kachru$^2$, and Arnav Tripathy$^3$}
\bigskip
\bigskip
\centerline{$^1$Institute for Gravitation and the Cosmos}
\centerline{The Pennsylvania State University, University Park, PA 16802}
\medskip
\centerline{$^2$Stanford Institute for Theoretical Physics}
\centerline{Stanford University, Stanford, CA 94305}
\medskip
\centerline{$^3$Department of Mathematics}
\centerline{Harvard University, Cambridge, MA 02138}

\bigskip
\bigskip
\begin{abstract}
Attractor black holes in type II string compactifications on $K3 \times T^2$ are in correspondence with equivalence classes of binary quadratic forms. The discriminant of the quadratic form governs the black hole entropy, and the count of attractor black holes at a given entropy is given by a class number. Here, we show this tantalizing relationship between attractors and arithmetic can be generalized to a rich family, connecting black holes in supergravity and string models with analogous equivalence classes of more general forms under the action of arithmetic groups. Many of the physical theories involved have played an earlier role in the study of ``magical" supergravities, while their mathematical counterparts are directly related to geometry-of-numbers examples in the work of Bhargava et. al.

This paper is dedicated to the memory of Peter Freund. The last section is devoted to some of  M.G's personal reminiscences of Peter Freund.

\end{abstract}

\section{Introduction}


Studies of BPS states provide one of the few windows we have into the structure of non-perturbative field theory and string theory.  It is therefore
interesting to ask if underlying (perhaps non-manifest) mathematical structures can be uncovered from studies of BPS spectra.
One set of hints in this direction appears in the papers  \cite{Mooreone,Mooretwo}, where connections between arithmetic and the study of attractor black holes \cite{attractors} were discussed.
A particularly striking observation in these papers relates
the numbers of attractor black holes at a fixed entropy in
$K3 \times T^2$ compactification of type II strings to
class numbers of binary quadratic forms with negative
discriminant.

These numbers are defined as follows.  Consider
quadratic forms
$$ax^2 + b xy + c y^2~,$$ which have integral coefficients and are positive definite, so that in particular the discriminant $$D = b^2 - 4ac$$ is negative. The arithmetic group $SL(2, \mathbb{Z})$ acts on such quadratic forms by simply acting on $(x, y)$ by the $SL(2, \mathbb{Z})$-transformation. The number $H(D)$ of such $SL(2, \mathbb{Z})$-equivalence classes at a fixed discriminant $D$ defines the (Hurwitz) class number, which in turn may immediately be related to class numbers of imaginary quadratic fields. For example, if $D$ is squarefree, $H(D)$ is simply the class number\footnote{Strictly speaking, the Hurwitz class number also has some fractional contributions to correctly account for automorphisms.} of $\mathbb{Q}[\sqrt{D}]$.

The low energy effective supergravity describing the $K3 \times T^2$ compactification of type II superstring belongs to the family of  $N=4$ Maxwell-Einstein supergravity theories that describes the coupling of  $N=4$ supergravity to   $N=4$ vector multiplets with the global symmetry group $SO(n,6) \times SU(1,1)$ -- realizing the specific case $n=22$.   Moore argues in \cite{Mooreone,Mooretwo} that given a BPS black hole with quantized electric and magnetic charges $(q, p)$ transforming in the $(n+6,2)$ representation of $SO(n,6) \times SU(1,1)$, one may associate the $SL(2, \mathbb{Z})$-equivalence class of binary quadratic forms with
$$a = {1\over 2}p^2,~b = p \cdot q,~c = {1\over 2}q^2~.$$
This association arises naturally from the geometry of the
attractor $K3 \times T^2$ associated to the black hole with
those charges.  Then $H(D)$ gives the number of distinct
attractor black holes at a fixed value of  the discriminant $D$, which coincides with the quartic invariant of $SO(n,6) \times SU(1,1)$
and also governs the supergravity black hole entropy:
$$S = \sqrt{p^2 q^2 - (p\cdot q)^2}~.$$
This relationship, and its extension to the class groups which endow the equivalence classes at
fixed $D$ with a group multiplication law \cite{Mooreone,Mooretwo}, was reviewed recently in \cite{Ono}.

It is natural to wonder whether this connection is an accident of special coincidences about the geometry of
$K3$ or if it might presage a more general set of relations between BPS black holes and structures in number theory.
Here, we give evidence for the latter viewpoint by generalizing this observation to a wider class of supergravity
theories, many of which arose originally in the study of ``magical'' supergravities in the early 1980s \cite{Muratone,
Murattwo}.
Many of these theories, in turn, admit embeddings into string theory.\footnote{
More precisely, we are only concerned with the geometry of the vector multiplet moduli space here.  So realizations of
these theories with extra purely neutral hypermultiplets will constitute UV completions of our picture.
In what follows, whether we know an explicit string
embedding or not, we will assume that the supergravity duality group is broken to a corresponding arithmetic group
in the full theory.  This is certainly what happens when a known string embedding exists.} In particular, we highlight the striking observation that the special points of arithmetic interest in many of Bhargava's geometry-of-numbers examples are in exact correspondence with special points from the point of view of supergravity via the attractor flow. 

The organization of this note is as follows.  In the next section, we describe examples of the physical
theories of interest.
In each, the classification of attactor points is equivalent to the problem of counting
equivalence classes of certain forms modulo the action of an arithmetic group; the black hole
entropy is then further controlled by a suitable arithmetic invariant.
In several cases, this corresponding algebraic structure
also appears in the study of Bhargava's geometry of numbers as originally studied in the context of higher composition laws \cite{Bhargava}, and the class numbers governing attractor degeneracies 
have been of independent interest in the study of arithmetic statistics.\footnote{For an accessible introduction to this
subject, see e.g. \cite{statistics}.}  We conclude by discussing some natural
directions for future exploration.  In an appendix we describe in more detail the connection of the structures we study to Jordan
algebras.

\section{Physical theories and mathematical structures}

In this section, we describe in telegraphic terms some of the
supergravity theories (often with string embedding) which
figure in the sequel.  They give rise to mathematical structures which can be related to the work of
Bhargava, as discussed in a different language in e.g. the work of Krutelevich \cite{Krut}; see particularly Table 1 there.
We should stress that Krutelevich uses the language of Jordan algebras and related Freudenthal triple systems which underlie the symmetry groups and geometries of all extended Maxwell-Einstein supergravity theories with symmetric target spaces in d$=5,4,3$, in particular the magical supergravity theories \cite{Muratone},  as well as $N > 4$ simple supergravity theories. For a review and references on the subject we refer to \cite{Gunaydin:2009pk}.

\subsection{Example one}

Consider pure 5d $N=2$ Poincare supergravity, or more generally a 5d $N=2$ theory with no vector multiplets (but perhaps coupled to hypermultiplets), a theory with eight supercharges.  Upon
compactification on a circle, one obtains 4d $N=2$ supergravity
with a single  vector multiplet (coming from Kaluza-Klein
reduction of the graviton on the circle, plus its superpartners) plus, perhaps, neutral hypers.\footnote{ We should note that there is another d=4 $N=2$ Maxwell-Einstein supergravity with one vector multiplet which does not have an uplift to five dimensions. Its quartic invariant is given by the square of a quadratic form.} The cubic form that defines this theory is simply $\mathcal{N}= X^3$ which leads to the prepotential $${\cal F} = X^3/X_0$$
in four dimensions.

The black holes in this theory are characterized by two electric charges (under the matter vector multiplet and the
graviphoton), and the corresponding magnetic charges.  They transform in the spin $s=3/2$ representation of an $SL(2,{\mathbb R}$)
duality group of the 4d supergravity, as discussed in~\cite{MizoguchiOhta}. 
  We should note that the attractor flows for BPS black holes in general homogeneous  supergravity theories, including this theory, were studied in \cite{Gunaydin:2005mx,Gunaydin:2007bg}. The attractor flows for non-supersymmetric  black holes in this theory were studied in detail in \cite{Gaiotto}.

In a non-perturbative completion of this theory, one expects $SL(2,{\mathbb R})$ to be demoted to $SL(2,{\mathbb Z})$.
There is a U-duality invariant formula for the black hole entropy in such a theory, given in e.g. \cite{Muratpure}.
We find it to be in beautiful correspondence with the following
construction.

Consider a binary cubic form
\begin{equation} F(x,y) = ax^3 + bx^2y + cxy^2 + dy^3~. \label{cubic}\end{equation}
It has a discriminant given by
\begin{equation} D = 18 abcd + b^2 c^2 - 4ac^3 - 4b^3d - 27a^2d^2~.\label{DiscD}\end{equation}
By a simple change of variables,
\[ a=-\xi_0 /3 , \quad b= \xi_1/3 , \quad c= - \eta_1/3 , \quad  d= \eta_0/3 \]
it takes the form
\[ D= -\tilde{I}_4 /3 \]
where
\[ \tilde{I}_4= (\xi_0)^2 (\eta_0)^2 + 4 (\xi_1)^3 (\eta_0) + 2 \xi_0 \eta_0 \xi_1 \eta_1 -1/3 (\xi_1)^2 (\eta_1)^2 - 4/27 (\xi_0) (\eta_1)^3 \]
is the $SL(2,\mathbb{R})$ invariant quartic form corresponding to
 eqn. (3.40) of \cite{Muratpure}.  This in turn governs
the entropy of 4d charged black holes in our supergravity theory.

There is a natural action of $SL(2,{\mathbb Z})$ on $(x,y)$ which
induces an action of $SL(2,{\mathbb Z})$ on the set of binary
cubic forms.  Defining equivalence classes as we did in the
case of binary quadratic forms, we obtain class numbers for
cubic forms $h_3(D)$.  These were studied by Davenport in 1951
\cite{Davenport}.  These count the inequivalent attractor
black holes (at fixed entropy)
in the physical theory we are describing.
The promotion of these class numbers to orders of class groups
is described in \cite{Bhargava}.

The geometry of numbers of this example proceeds by an intriguing quadratic map from binary cubic forms to binary quadratic forms originally investigated by Eisenstein \cite{Eisenstein}.  We refer to \cite{Hoffman} for a review of Eisenstein's work and further references on the subject, which we explain here to indicate the special point structure as well as the fundamental importance of the quadratic transformation of the charge lattice.

Indeed, given a binary cubic form $F(x,y)$ as in equation~\eqref{cubic} with discriminant $D$ one can associate with it a closely-related binary quadratic form $Q_F(x,y)$ essentially given by the Hessian of $F$, so that the association is naturally $SL(2, \mathbb{Z})$-equivariant. To be more explicit, it is convenient to restrict our cubic form slightly and instead take $$F(x, y) = ax^3 + 3bx^2y + 3cxy^2 + dy^3,$$ with $a, b, c, d$ still all integers. Then the associated quadratic
\begin{equation}  Q_F(x,y) = A x^2 +  B x y + C y^2 \end{equation} is given by \begin{equation}  A = b^2 - a c , \quad B= b c - a d , \quad C=  c^2 - bd  \end{equation} and the discriminant $D_Q$ of the quadratic form is related to the discriminant $D$  of the cubic form as
\begin{equation} D_Q = B^2 - 4 A C = -\frac{D}{27}. \label{FtoQ} \end{equation}

As mentioned, the mapping from $F(x,y)$ into  $Q_F(x,y)$ commutes with the $SL(2,\mathbb{Z})$ action, so we have a map from equivalence classes of integral binary cubic forms at a given discriminant to those of binary quadratic forms at (roughly) the same discriminant. It is obvious interest to understand if this map is one-to-one and to characterize its image. We mention here only the simplest case, of $D_Q \equiv 0\pmod{4}$ and $D_Q/4$ squarefree, in which case the map is indeed injective but the image is striking: one obtains exactly those points of order $3$ in the class group $\mathrm{Cl}(\mathbb{Q}(\sqrt{D_Q}))$.

Of particular interest is the formula for the growth of the
class numbers $h_3(D)$ with $D$.  Davenport proves that\footnote{Note that the numerical coefficient of the first term on the right hand side   was corrected by Davenport to be $1/36$  in a corrigendum \cite{corrigendum}.  }
\be \sum_{N=1}^D h_3(N) = {\pi^2 \over 36} D + {\cal O}(D^{15\over 16}) \ee
  It would be interesting to reproduce this formula
directly from the perspective of supergravity counts of black
hole attractors.

Regarding a string construction of the pure supergravity theory with 8 supercharges
in five dimensions, its possible existence is discussed (as a ``fantasy island") in
\cite{islands} but more completely analyzed in~\cite{MizoguchiSchroder, Mizoguchi} as a consistent truncation of type II supergravity (while not directly arising as a massless sector of a string compactification). 

\subsection{Example two}

In a similar spirit, one can consider a theory of 4d $N=2$ supergravity coupled to two vector multiplets that descends from $N=2$ supergravity coupled to one vector multiplet in $d=5$. The cubic norm that defines this theory is $\mathcal{N}= X^2 Y$ which leads to the prepotential
$ X^2 Y/X_0$ in four dimensions. The scalar manifold of the $4d$  supergravity is the symmetric space
\[  \mathcal{M}_4 = \frac{SO(2,2)}{U(1)\times U(1)} \]
The discrete U-duality group is $SL(2,{\mathbb Z}) \times SL(2,{\mathbb Z})$  under which the six charges (including
electric and magnetic graviphoton charge)
transform in the $(3,2)$ representation.

The mathematical structure here is that of {\bf pairs} of binary quadratic forms, which are exchanged by one of the $SL(2,{\mathbb Z})$
symmetries.  The other acts on both (simultaneously) as described in the natural action on binary quadratic forms above.
A higher composition on pairs of binary quadratic forms is described in \cite{Bhargava}.

Given such a pair
$$ax^2 + b xy + cy^2$$
$$A x^2 + B xy + C y^2$$
there are three natural discriminants one can define; those of each of the quadratic forms $\Delta_{1,2}$, and the codiscriminant
$$\Delta_c = bB - 2aC - 2 cA~.$$
The problem of counting equivalence classes of such forms (up to action of the arithmetic group) for triples of these quantities
has been discussed by Morales \cite{Morales}.  This is a more refined count than is natural in supergravity, where the entropy would be determined
by the single duality invariant
$$\Delta_c^2 - 4 \Delta_1 \Delta_2~.$$
Hence, (sums of) the class numbers of Morales govern the attractor counts of black holes in this supergravity theory.

\subsection{Example three}

We can consider the famous STU model with three vector multiplets coupled to 4d $N=2$ supergravity.  This theory descends from the 5d Maxwell-Einstein supergravity with two vector multiplets defined by the cubic norm $ XYZ$ that was first studied in \cite{Gunaydin:1984ak}.
A nice discussion of BPS black holes in this model can be found in \cite{Kallosh}.
A promotion of this model
to string theory is described in \cite{SenVafa} (where it is basically $N=2$ Example D) and in \cite{Kounnas}.\footnote{It
is also easy to construct Calabi-Yau threefolds with $h^{2,1}=3$ and a complex structure moduli space governed by a prepotential of this form; we thank J. Bryan for discussions of this point.}
  The prepotential in d$=4$ is $XYZ/X_0$ and is generally denoted as
$$F = STU~$$
in the gauge $X_0=1$.
The model has an $SL(2,{\mathbb Z})^3$ duality symmetry.\footnote{The $SL(2,{\mathbb Z})$ factors are broken slightly to congruence subgroups in
the known string embeddings, which would cause minor
modification to the discussion below.}
The 8 electric and magnetic charges transform in the
$(2,2,2)$ representation of the duality group.

The algebraic structure involved here is
$$V_2 \otimes V_2 \otimes V_2$$
with $V_2$ the two dimensional representation of
$SL(2,{\mathbb Z})$.  Explicit actions of the duality
group on electric and magnetic charges can be found
in \cite{Kallosh}.  In a suitable
basis of the electric and magnetic
charges, the duality invariant
entropy is again given by
$$S = \sqrt{p^2 q^2 - (p \cdot q)^2}~.$$
Again, class numbers are determined by the numbers of
elements in $V_2^{\otimes 3}$ modulo the action of
$SL(2,{\mathbb Z})^3$, and give the numbers of distinct attractor
points at different values of the entropy.

\subsection{Example four}

There is a similar story relating counts of BPS black holes to
binary quartic forms.  Consider the family of forms
$$f(x,y) = a x^4 + b x^3y + c x^2 y^2 + d xy^3 + e y^4~.$$
Again, there is a natural $SL(2,{\mathbb Z})$ symmetry that acts on $(x,y)$ and induces an action on the space of forms.
There are now two invariants:
$$I(f) = 12 ae - 3bd + c^2$$
$$J(f) = 72 ace + 9 bcd - 27 ad^2 - 27 e b^2 - 2 c^3~.$$
The discriminant of $f$ is
$$\Delta(f) = {1\over 27} \left( 4I(f)^3 - J(f)^2 \right)~.$$

The discriminant $\Delta(f)$ has the form of the entropy of the 5d uplift of an extremal black hole of $N=2$ Maxwell-Einstein supergravity with one vector multiplet in d=4, except for the fact that it involves an extra parameter in addition to four charges (two electric and two magnetic) with respect to the 4d vector fields. This is easily established by setting the integer $e$ in the binary cubic form equal to zero. We then find that the discriminant takes the form
\[ \Delta(f)|_{e=0}  = d^2 D \]
where $D$ is the discriminant of the binary  cubic form  given above.  The quartic invariant  $I_4$ that defines the entropy of charged black holes given in equation (3.40) of \cite{Muratone} can be written in a more symmetrical  way between the electric and magnetic charges, via the identification
\[ \xi_0= - q_0=- 9a ,\quad \eta_0 = p_0=d/3, \quad \xi_1 = q_1/3=b/3 , \quad \eta_1=3 p_1=c \]
Then the quartic invariant takes the form
\[ I_4 =\tilde{I}_4 / 9= - D/3 = 4 (p_1)^3 q_0 +4 (q_1)^3 p_0 + (p_0)^2 (q_0)^2-6 p_0 q_0 p_1 q_1 - 3( q_1)^2 (p_1)^2 \]
 The quadratic and cubic invariants $I(f)$ and $J(f) $ associated with the binary quartic form then take the following form in terms of  these variables and the extra parameter $e$
 \[ I(f)= -4/3 e q_0 -9 p_0 q_1 +9 (p_1)^2 \]
 \[ J(f) =3 \left( -9 e (q_1)^2 -9  q_0 (p_0)^2 +8  e q_0 p_1 + 27 p_1 p_0 q_1 -18 (p_1)^3\right) \]

 Now for $e=0$ the discriminant $\Delta$ describes the duality invariant form governing the entropy of a 5d uplift of an extremal black hole of a 4d, $N=2$ Maxwell-Einstein supergravity with one vector multiplet, since
 \begin{equation} \Delta|_{e=0} = - 27 (p_0)^2 I_4 ~.
 \end{equation}

 This follows from the general result that the entropy of a spinning charged black hole (or ring) that is an uplift of 4d extremal black hole with charges $p^0, p^i, q_0, q_i$ has the form \cite{Gaiotto:2005xt,Gaiotto:2005gf,Pioline:2005vi}  $$S_{5d} = (2\pi) \sqrt{\left( N_3(Q_i) - J^2 \right)} ~.$$ where

$$Q_i = p^0 q_i + {1\over 2} C_{ijk}p^jp^k$$
$$N_3(Q_i) = {1\over 6} C^{ijk} Q_i Q_j Q_k$$
$$J = {1\over 2}\left( p^0 (p^0 q_0 + p^i q_i) + {1\over 3}
C_{ijk} p^i p^j p^k\right)~.$$
Then the entropy of the corresponding 4d BPS black hole  is given by
$$S_{4d} = {2\pi \over |p_0|} \sqrt{N_3(Q_i) - J^2},$$
where $J$ is the 5d angular momentum.

Therefore the invariant $J(f)|_{e=0}$  describes the 5d angular momentum of the spinning black hole in our example.
We should note that the
general black hole solution of pure  $N=2$ supergravity in five dimensions involving six parameters, namely the four electromagnetic charges, mass and angular momentum was given in \cite{Tomizawa:2012nk} to which we refer for references on earlier work on the subject. However the general  solution has not been written in terms of an invariant corresponding to the discriminant $\Delta$ with non-vanishing parameter $e$. On the basis of the above analysis we predict that there exists a five parameter family of extremal black ring solutions  of pure $N=2$ 5d supergravity whose entropy is described by the invariants  $J(f), I(f)$ and $\Delta(f)$.

 As was shown in \cite{quartic} for fixed invariants $I(f)$ and $J(f)$ there exists  a single orbit of $SL(2,\mathbb{R})$ in the space of quartic forms $F(x,y)$ if $\Delta(f) <0$ and the orbit lies in the subspace of quartic forms with one pair of complex roots. For fixed
 invariants $I(f)$ and $J(f)$  with positive discriminant $\Delta(f)>0$ one finds three different orbits of  $SL(2,\mathbb{R})$. Two of these orbits lie in the subspace where the quartic form admits two pairs of complex roots while the third orbit lies in the subspace with real roots only.

Aspects of the asymptotics of class numbers $h(I,J)$ of binary quartic forms
with given values of $I, J$ are determined by Bhargava and Shankar in \cite{quartic}.  Theorem 1.6 of that paper tells
us the following.  If we define
$$H(I,J) = {\rm max}{|I^3|,|J^2/4|}~,$$
then
$$\sum_{H(I,J) < X} h(I,J) \sim \zeta(2) X^{5/6} +
{\cal O}(X^{3/4 + \epsilon})~.$$
Again, it would be nice to present a physics proof.
\subsection{Magical Supergravities}
In this section we will give an example related to one of the magical supergravity theories \cite{Muratone,Murattwo} and indicate how this example could  be extended to the other magical supergravity theories, that are briefly reviewed in the Appendix.

Consider the extension of example 1 to the case of ternary quadratic forms\cite{Gross_Lucianovic}
\be
Q(x,y,z) = a x^2 + b y^2 + c z^2 + u yz + v xz + w xy  \label{Quat}
\ee
The (half)-discriminant of  $Q(x,y,z)$  is given by
\be
\Delta_Q =\frac{1}{2} Z =  4 abc + uvw -a u^2 -b v^2 - c w^2
\ee
where
\be
Z = \begin{pmatrix}
      2a & w & v \\
      w & 2b & u \\
      v& u & 2c \\
    \end{pmatrix}
\ee
Therefore the matrix $Z$ can be identified with the charges of extremal black hole solutions in real magical supergravity defined by the Jordan algebra $J_3^{\mathbb{R}} $ of $3\times 3$  symmetric matrices and the discriminant $\Delta_Q $ determines their entropy.  Under the action of
 $5d$ U-duality group $SL(3,\mathbb{R})$ of real magical supergravity on $(x,y,z)$ the ternary quadratic form transforms as follows:
\be
g : \left(
      \begin{array}{c}
        x \\
        y \\
        z \\
      \end{array}
    \right) \longrightarrow  \left( \begin{array}{c}
                                 x'\\
                                 y' \\
                                 z' \\
                               \end{array} \right)=g^{-1} \centerdot \left(
      \begin{array}{c}
        x \\
        y \\
        z \\
      \end{array}\right)  \Longrightarrow
Q(x',y',z')=g\centerdot Q(x,y,z)
\ee
where $g\in SL(3,\mathbb{R})$.  One finds that the discriminant , that determines the entropy, is left invariant under this action of $SL(3,\mathbb{R})$:
\be
\Delta[Q(x',y',z')] =Det(g)  \Delta[Q(x,y,z)] = \Delta[Q(x,y,z)]
\ee

The prepotential of the real magical supergravity in four dimensions is given by
$${\cal F} \sim i {{\rm det}(X) \over X^0}~,$$
where $X\in J_3^{\mathbb{R}}$.  The U-duality group of the four dimensional supergravity is  $Sp(6,\mathbb{R})$ with the maximal compact subgroup $U(3)$. The entropy of $4d$ BPS black hole solutions is given by the quartic invariant of the Freudenthal triple system associated with $J_3^{\mathbb{R}}$\cite{Ferrara:1997uz,Gunaydin:2000xr,Gunaydin:2009pk}.
This vector multiplet moduli space is $Sp(6,\mathbb{R})/U(3)$ and corresponds to the item 9  in Table 1
of \cite{Krut}. The charges transform in the 14 dimensional
irreducible representation of the arithmetic group, and the
relevant  highest weight module $V(\omega_3)$ is identified in
\cite{Krut} and serves as the analogue of the binary forms in our previous examples.  The Fourier coefficients of modular forms on $PGSp(6)$ and their connection to quaternion rings  were studied in \cite{Lucianovic}.

We should note that both Krutelevich and Bhargava in the cited work \cite{Krut} restrict themselves to split real forms of the corresponding U-duality groups\footnote{We would like to thank one of the referees of our paper for reminding us of the fact that the symmetry groups appearing in the work of Bhargava and Krutelevich  are all of the split real form. This prompted us to replace, as our main  example, the complex magical supergravity theory of the  first submission with the example of  the real  magical supergravity. We had discovered the connection between the real magical supergravity and ternary quadratic forms  after the first submission to the arXiv and were originally contemplating writing  it up as a separate publication.}. For the real magical supergravity the U-duality groups in five , four and three dimensions are all of the split form. However the case of real magical supergravity does not appear in Bhargava's work and Krutelevich suggests that it could produce a new example of a space with higher composition law. Here we identify the corresponding supergravity theory as the real magical supergravity theory which can be obtained by orbifolding the quaternionic magical supergravity\cite{Muratone} constructed by Sen and Vafa from superstring theory by orbifolding using the dual pair method\cite{SenVafa}.

In Krutelevich's table the rows 5,6 and 7 correspond to  Jordan algebras of $3\times 3 $ Hermitian matrices over the split composition algebras with associated symmetry groups $SL(6,\mathbb{R})$, $SO(6,6)$ and $E_{7(7)}$. Of these only the one corresponding to split complex numbers appears in Bhargava's work. Its symmetry group is  $SL(6,\mathbb{R})$.  The split exceptional Jordan algebra is associated with the maximal $N=8$ supergravity with its $4d$ U-duality group $E_{7(7)}$. Maximal supergravity can be truncated to a purely bosonic theory with $SO(6,6)$ symmetry or $SL(6,R)$ symmetry . Since they are not supersymmetric one can not have BPS black holes in these truncated theories.

On the other hand maximal supergravity can be truncated to quaternionic magical supergravity with $SO^*(12)$ symmetry and 15 vector multiplets with target space $SO^*(12)/U(6)$  in four dimensions. This is the theory obtained by Sen and Vafa from superstring theory by orbifolding.
Quaternionic theory can further be truncated to the complex magical supergravity with $SU(3,3)$ symmetry and 9 vector multiplets\cite{Muratone,Murattwo}.
$${\cal M}_{\rm vector} = SU(3,3;{\mathbb Z}) \backslash SU(3,3) \slash (SU(3) \times SU(3) \times U(1))~.$$
The moduli space is 9 (complex) dimensional, and the prepotential
is given by
$${\cal F} \sim i {{\rm det}(X) \over X^0}~,$$
where $X$ is a three-by-three Hermitian matrix. It descends from the complex  magical supergravity in $5d$ with the scalar manifold  $$ {\cal M}_5 = SL(3,\mathbb{C})/SU(3). $$
 This model was known to have a string theory embedding, as a ${\mathbb Z}_3$ orbifold of $T^6$ compactification before the work of Sen and Vafa and was discussed in
in detail in \cite{Ferrara}.

Underlying Jordan algebras of  magical supergravity theories are all Euclidean defined by  $3\times 3$ Hermitian matrices over the four division algebras. This and the above mentioned facts suggest  that Krutelevich's work and related Bhargava's higher composition laws may be extended along magical lines by replacing the split Jordan  algebras with the Euclidean Jordan algebras. Their symmetry groups will be different real forms of  the split real forms appearing in Krutelevich's table including the exceptional Jordan algebra which underlies the octonionic magical supergravity with target space $E_{7(-25)}/E_6\times U(1)$ in four dimensions.

We should also stress the important fact for all the rows in Krutelevich's table  that have a supergravity realization the last column lists symmetry groups that are isomorphic to the isometry groups of the three  dimensional supergravity theories obtained by dimensional reduction. These groups have been proposed as spectrum generating symmetry groups of $4d$ black holes\footnote{See the lectures \cite{Gunaydin:2009pk} and the references therein.}. Furthermore the Fourier coefficients of modular forms associated with the minimal unitary representations of the three dimensional U-duality groups are expected  to describe the degeneracies of $4d$ black    holes\cite{Pioline:2005vi,Muratpure,Gunaydin:2005mx}. For the examples discussed above these groups are $G_{2(2)}, SO(4,3), SO(4,4)$ and $F_{4(4)}$ corresponding to $d=4$ $N=2$ supergravity coupled to  one, two, three and  six vector multiplets, respectively.


\section{Discussion}

In this note, we outlined a connection between special points in geometry-of-numbers examples and special points in supergravity. If one wishes to count these points, one matches class numbers characterizing forms with fixed values of various arithmetic invariants and BPS black holes in supergravity and string theory. 
While we provided a few examples in \S2, it should be
clear that similar considerations extend to many further constructions.
Extension of our results to other supergravity theories, in particular those that have stringy extensions, and possible extensions along the lines of example four above will be left to future investigations.


In the prototype case involving binary quadratic forms and BPS black holes on $K3 \times T^2$,
it has been observed that the generating function
$$\sum_{D} H(D) ~q^D$$
(with suitable constant term)
provides the holomorphic piece of a weight 3/2 mock modular form \cite{Tripathy}.
It would be very interesting to relate the black hole counting problems above to
modular objects in a similar manner.  In fact, the general statements of Kudla-Millson theory, as described in \cite{KT}, encapsulate the count of attractor black holes in our Example 2
in a degree 2 weight 2 Siegel form studied by Kohnen \cite{Kohnen}. 
Kohnen's $Sp(4,\mathbb{R})=Spin(3,2)$ symmetry is a subgroup of the U-duality group $SO(4,3)$ of the corresponding three dimensional supergravity theory that was proposed as spectrum generating symmetry group of the $4d$ supergravity as mentioned above. 

The results here suggest many other natural questions:

\medskip
\noindent
$\bullet$ Can we find proofs of the asymptotic results governing class numbers (some of which were presented above) using
physical arguments?

\medskip
\noindent
$\bullet$ The theories most closely tied to these arithmetic variants seem to be $N\geq 2$ supersymmetric theories which enjoy a certain finiteness property (no instanton corrections to the prepotential).  Can we demonstrate a connection of all such $N=2$ theories to the theory of arithmetic invariants?  Can we turn this around and propose a classification of such theories in terms of invariant theory?

\medskip
\noindent
$\bullet$ Most ambitiously, one would like to find a version of this story which holds in quantum geometry, e.g. for counts of BPS black holes in the mirror quintic.  This should involve a suitable generalization of the various classical number theoretic notions that played a role here.  Progress on this front would be most interesting.

 \section{Remembering Peter Freund ( M.G.) }
 Scientifically I got to be introduced to  Peter Freund for the first time when I received a preprint of his paper on " Quark parastatistics and color gauging"  in the mid 1970s in which he coined the term GG statistics referring to my work with my advisor Feza G\"ursey on the connection between color quarks and octonions.
 Personally I got to meet him for the first time in a workshop at the University of Washington organized by Tony Zee at the beginning of 1980s. The participants of that workshop were all housed in a sorority building on the campus of the university. One of the most memorable events of the workshop was the five course dinner  that John Iliopoulos cooked, in the kitchen of the sorority, for all the participants which was certainly worthy of some Michelin stars. The evening was capped  by Peter singing the Kindertotenlieder, the song cycle  of Mahler, in the accompaniment of another physicist at the piano. It was an emotional performance. Afterwards Peter told me that singing the Kindertotenlieder moves him always very deeply.

 My next encounter with Peter was at the University of Chicago where I  gave a seminar  on my work on magical supergravity theories which found a very receptive audience in him. I appreciated greatly the stimulating discussions I had  with him after my talk.

 Peter's work on  Kaluza-Klein supergravity, in particular his paper with Rubin on $S^7$ compactification of 11 dimensional supergravity, had a major impact in the field including some of my own work.   My work on  the construction of the spectra   IIB supergravity on $S^5$ ( with Marcus), of 11d supergravity on $S^7$ ( with Warner ) and on $S^4$ ( with van Nieuwenhuizen and Warner),by a simple tensoring procedure from some fundamental supermultiplets,  represent some of the earliest work on AdS/CFT dualities within the framework of Kaluza-Klein supergravity in a true Wignerian sense. All this work was done during the summer of 1984 in Aspen. Next door to us Mike Green and John  Schwarz were busy working on their famous anomaly cancellation mechanism in superstring theory. I do not recall seeing  Peter in Aspen that summer. Maybe we simply did not overlap.

Peter attended a conference in memory of my advisor Feza G\"ursey, who was his friend, that we organized on the campus of my alma mater, Bogazici University ( formerly Robert College) during the summer of 1994. He clearly enjoyed his visit to Istanbul, the conference  and the historical sites.  It was a pleasure to be in his company. He always emanated a sense of enthusiasm and excitement  whatever he talked about with his baritone voice.

Last time I saw  him was at  Gunnar Nordstr\"om Symposium on Theoretical Physics at the University of Helsinki in  Finland on August 28, 2003.

Peter had very broad interests in physics as well as outside physics and has made many important original contributions to theoretical physics. Outside of his work on Kaluza-Klein supergravity he is probably best known for his work on p-adic strings.

Peter Freund  belonged to that endangered species of scholarly gentleman physicists of the Old World. I will always remember him fondly.

\vspace{2cm} 

{\bf  Note added:}
After we submitted the published version of our paper to the arxiv   the authors of \cite{Borstenetal} brought their work to our attention which appeared after our first version and before the second version which we submitted to arXiv after publication. Their work deals mainly with the split real forms such as the maximal supergravity and its truncations to SO(6,6) and SL(6,R) invariant sectors and their relation to higher composition laws. We did not consider these truncations since they are not supersymmetric.  They also  point out that the complex magical supergravity example does not correspond to any of Bhargava's composition laws and add that it does not mean that such a composition law does not exist. They also indicate  potential problems in relating the  octonionic magical supergravity theory to  Bhargava's higher composition laws.

We believe that the example of real magical supergravity and its relation to ternary quadratic forms which we gave in our published version and the arguments presented therein suggest  that there must exist  extensions or generalizations of Bhargava's composition laws to novel higher composition laws related to the magical supergravity theories based on simple Euclidean Jordan algebras of degree three.

The authors of \cite{Borstenetal} also study the orbits of the STU model under its discrete U-duality group $SL(2,\mathbb{Z})^3$  using Bhargava's cube. We refer to  \cite{Borstenetal} for details and the references to their earlier work on the application of Bhargava's cube  to STU model. They state that their discussion of STU orbits  has significant overlap with the results of \cite{Senetal} which appeared shortly before their work. The authors of \cite{Senetal} also use the method of  Bhargava's cube   and  give a more complete classification of the orbits of STU models under $SL(2,\mathbb{Z})^3$ as well as its congruence subgroups.

\bigskip
\centerline{\bf{Acknowledgements}}
\bigskip

We would like to thank Greg Moore for fun discussions.
M.G. and S.K. thank the Simons Foundation for hospitality during
the closing stages of this work.
M.G. would also like to thank Manjul Bhargava for stimulating discussions on his work on higher composition laws and their possible connections to physics during his visit to deliver the  2011 Marker Lectures in Mathematics at Penn State University.
He acknowledges gratefully the hospitality of the Stanford
Institute for Theoretical Physics, where part of this work was
done.  S.K. thanks J. Bryan for some quick discussions of Calabi-Yau moduli spaces that might be of interest for this program.
He was supported in part by
the National Science Foundation under grant PHY-1720397, and by a Simons Investigator Award. A.T. would like to thank Arul Shankar for his inspiring comments and patient insight, as well as the support of the NSF under grant 1705008.

\section*{Appendix}

The examples we considered in this paper correspond to the $N=2$ Maxwell-Einstein supergravity theories with zero, one, two and five vector multiplets in d=5 and their 4d counterparts. Five dimensional $N=2$ Maxwell-Einstein supergravities with symmetric target spaces $G/H$,such that $G$ is a symmetry of the Lagrangian, are in one-to-one correspondence with the Euclidean Jordan algebras of degree three\cite{ Muratone,Murattwo}. The $C$-tensor $C_{IJK}$ appearing in the coupling
$$C_{IJK} F^I \wedge F^J \wedge A^K$$ in these theories
 determines their Lagrangian uniquely and is given by the cubic norm of the underlying Jordan algebra $J$. The Jordan algebras of degree three admit 3 idempotents $e_1, e_2$ and $e_3$ with the identity element given by
$$\mathbb{I}= e_1 + e_2 + e_3$$
 which corresponds to the bare graviphoton in the corresponding supergravity theory. Hence the pure $N=2$ supergravity is described by the norm
$$\mathcal{N}_3 (X \mathbb{I})= X^3~.$$
The example 2 we considered above corresponds to the $N=2$ Maxwell-Einstein supergravity described by the cubic norm
$$\mathcal{N}_3 ( X e_1 + Y e_2 + Y e_3) = X Y^2 $$
and the STU model in five dimensions is defined by
$$\mathcal{N}_3 ( X e_1 + Y e_2 + Z e_3 )= XYZ$$
as found in \cite{Gunaydin:1984ak}. The real magical supergravity in $5d$ is defined by the Jordan algebra $J_3^{\mathbb{R}}$ of three-by-three real symmetric  matrices and the norm of an element $J$ is given by its determinant
$$ \mathcal{N}_3(J)= det(J)$$
 By the action of the automorphism group $H$ of the underlying Jordan algebra every element of  $J$   can be brought to the form  $( X e_1 + Y e_2 + Z e_3 )$ of the STU model. For the real magical supergravity the automorphism group of $J_3^{\mathbb{R}}$ is $SO(3)$. This shows clearly that there  exist  natural extensions of the invariants we studied  to more general classes of invariants related  to the arithmetic subgroups of the global symmetry groups of the $N=2$ supergravity theories. For $N=4$ Maxwell-Einstein supergravities in 5d, the underlying Jordan algebras of degree three  are non-Euclidean. Similarly the symmetries of maximal $N=8$ supergravity are given by the symmetries of the non-Euclidean  exceptional Jordan algebra of Hermitian  $3\times 3 $ matrices over the split octonions\cite{Ferrara:1997uz}. Under dimensional reduction to four dimensions the correspondence between Jordan algebras of degree three  and 5d supergravities  goes over to a  correspondence between Freudenthal triple systems $\mathcal{F}(J)$ associated with the Jordan algebras $J$ of degree three and 4d supergravities \cite{Ferrara:1997uz,Gunaydin:2000xr,Gunaydin:2009pk}.

 Orbits of extremal  black hole solutions of supergravity theories under the action their continuous U-duality groups $G$ have been studied extensively beginning with the work of \cite{Ferrara:1997uz}. For those theories that have stringy extensions the relevant orbits are with respect to the discrete  U-duality groups which are typically the maximal arithmetic subgroups $G(\mathbb{Z})$ of $G$ \cite{Maldacena:1999bp,Benjamin:2017xen}.
Motivated by the works of \cite{Ferrara:1997uz,Maldacena:1999bp} Krutelevich considered  the integral version of Freudenthal's construction of exceptional groups and  studied their connection to higher composition laws of Bhargava\cite{Krut}.    Of all the examples discussed in this paper only the example 4 does not appear in Krutelevich's table.

  \end{document}